# An Adaptive Watermarking Technique for the copyright of digital images and Digital Image Protection

Yusuf Perwej[1], Firoj Parwej[2], Asif Perwej[3]

[1]M.Tech, MCA, Department of Computer Science & Information System,
Jazan University, Jazan, Kingdom of Saudi Arabia (KSA)
Yusufperwej@gmail.com

[2]MCA, MBA, BIT, Department of Computer Science & Information System, Jazan
University, Jazan , Kingdom of Saudia Arabia(KSA)
firojparwej@gmail.com

[3]P.hD, MBA, Assistant Professor, Senior Member of the International Association of
Computer Science and Information Technology (IACSIT), Singapore
asifperwej@gmail.com

## ABSTRACT

*The Internet as a whole does not use secure links, thus information in transit may be vulnerable to interruption as well. The important of reducing a chance of the information being detected during the transmission is being an issue in the real world now days. The Digital watermarking method provides for the quick and inexpensive distribution of digital information over the Internet. This method provides new ways of ensuring the sufficient protection of copyright holders in the intellectual property dispersion process. The property of digital watermarking images allows insertion of additional data in the image without altering the value of the image. This message is hidden in unused visual space in the image and stays below the human visible threshold for the image. Both seek to embed information inside a cover message with little or no degradation of the cover-object. In this paper investigate the following relevant concepts and terminology, history of watermarks and the properties of a watermarking system as well as a type of watermarking and applications. We are proposing edge detection using Gabor Filters. In this paper we are proposed least significant bit (LSB) substitution method to encrypt the message in the watermark image file. The benefits of the LSB are its simplicity to embed the bits of the message directly into the LSB plane of cover-image and many techniques using these methods. The LSB does not result in a human perceptible difference because the amplitude of the change is little therefore the human eye the resulting stego image will look identical to the cover image and this allows high perceptual transparency of the LSB. The spatial domain technique LSB substitution it would be use to use a pseudo-random number generator to determine the pixels to be used for embedding based on a given key. We are using DCT transform watermark algorithms based on robustness. The watermarking robustness have been calculated by the Peak Signal to Noise Ratio (PSNR) and Normalized cross correlation (NC) is used to quantify by the similarity between the real watermark and after extracting watermark.*

## Keywords

*Watermarking, Steganography, Peak Signal to Noise (PSNR), data hiding, Copyright Protection, Normalized Cross Correlation (NC).*

# 1. INTRODUCTION

In recent years it has been seen a rapid growth of network multimedia systems. This has led to an increasing awareness of how easy it is becoming to reproduce the data. The ease with which perfect copies can be made may lead to large-scale unauthorized copying, which is a great concern to the image, music, film, and book. Because of this concern over copyright issues, a number of technologies are being developed to protect against illegal copying. One of these techniques is the use of digital watermarks [1]. Digital watermarking focuses mainly on the protection of intellectual property rights and the authentication of digital media. The steganographic methods similar to digital watermarking methods hide information in digital media. Steganography differs from cryptography in the sense that where cryptography focuses on keeping the contents of a message secret, steganography focuses on keeping the existence of a message secret [2].

The main destination of steganography is to communicate securely in a completely undetectable manner and to avoid drawing suspicion to the transmission of a hidden data. It is not to keep others from knowing the hidden information, but it is also to keep others from thinking that the information even exists. If a steganography method causes someone to suspect the carrier medium, then the method has failed [3]. An ideal steganographic system would embed a large amount of information, perfectly secure with no visible degradation to the cover object. The example of an ideal watermarking system embeds an amount of information that could not be erased and altered without making the cover image completely unusable. In this paper we are discussing LSB method and which will be applied in the frequency domain by selecting the pixel based on the frequency. We will begin with a quick background on steganography, which forms the basis for a large number of digital watermarking concepts.

# 2. INFORMATION HIDING

There are several techniques for information hiding in digital media. The basic method of information hiding is steganography. The term steganography know as cover writing. The growing possibilities of modern communications need the special means of security especially on computer network. The network security is becoming more important as the number of data being exchanged on the Internet increases. Therefore, the confidentiality and data integrity are required to protect against unauthorized access and use. This has resulted in an explosive growth of the field of information hiding. The rapid growth of publishing and broadcasting technology also require an alternative solution in hiding information. The copyright such as audio, video and other source available in digital form may lead to large scale unauthorized copying. This is because the digital formats make possible to provide high image quality even under multi-copying. Therefore, the special part of invisible information is fixed in every image that could not be easily extracted without specialized technique saving image quality simultaneously.

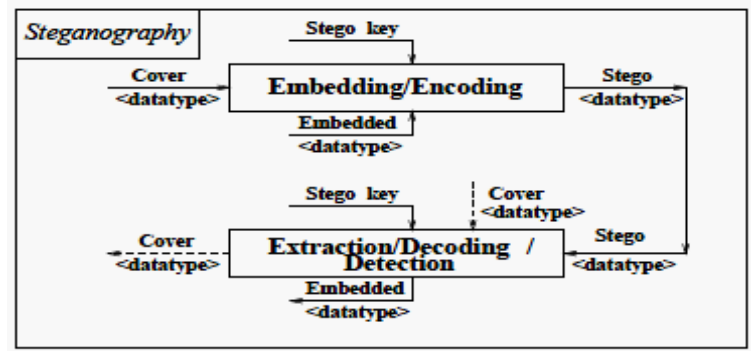

Figure – 1   Steganography

All this is of great concern to the music, film, book and software publishing industries. Information hiding is an emerging research area, which encompasses applications such as copyright protection for digital media, watermarking, fingerprinting, and steganography. Steganography is the art of secret communication [5]. Figure 1 shows the Steganography is a new method of making the communication more secured. These techniques gained importance in many diverse areas. Steganography is a storage mechanism designed to give user highest level of security against being compelled to disclose its contents [6]. It will deliver a file to any user who knows the name and password. Anyone else who does not possess this information can not gain any clue about the file hidden. It literally means "covered writing". The hidden messages could be plain text, cipher text, or anything that can be represented as a bit stream. These messages are put into carrier media such as innocent images, audio, video, text, or any other digitally represented code.

## 3.  DIGITAL WATERMARKING

Watermarking of papers was first done in Fabriano, Italy in 1292 [7]. During manufacturing a water coated stamp was impressed on the still not dry paper pulp. The watermark in the paper was visible as a lighter pattern when the paper was viewed by transmitted light. The watermark was used by paper makers to identify their products. Also referred to as simply watermarking a pattern of bits inserted into a digital image, audio or video file that identifies the file's copyright information. The name comes from the faintly visible watermarks imprinted on stationery that identify the manufacturer of the stationery. The purpose of digital watermarks is to provide copyright protection for intellectual property that's in digital format. It can also contain device control code that prevents illegal recording. An application of watermarking is copyright control, in which an image owner seeks to prevent illegal copying of the image. There is no evidence that watermarking techniques can achieve the ultimate goal to retrieve [8] the right owner information from the received data after all kinds of content-preserving manipulations. Because of the fidelity constraint, watermarks can only be embedded in a limited space in the multimedia data. There is always a biased advantage for the attacker whose target is only to get rid of the watermarks by exploiting various manipulations in the finite watermarking embedding space [9]. A more reasonable expectation of applying watermarks techniques for copyright protection may be to consider specific application scenarios.

## 4. TYPES OF WATERMARKS

Basically the watermarking is a two type that is embedded into a media object can either be perceptually visible or invisible.

## 4.1 Visible Watermarks

A visible watermark is a visible translucent image that is overlaid on the primary image. Visible watermarks change the signal altogether such that the watermarked signal is totally different from the actual signal, for example, adding an image as [9] a watermark to another image. A visible watermark can be seen by the human eye and often it contains a name or logo of a company or copyright information. This type of mark is intended to be easy to read by a receiver and deter pirate distribution since everyone can see the origin of the video. The visible watermark makes it easier for the pirate to see if an attack was successful. The example in the figure 2 shows both a watermark and an image with the overlaid watermark.

## 4.2 Invisible Watermarks

Invisible watermarks a copy should be indistinguishable from the original, i.e. the embedding of the watermark should not introduce perceptual distortion of the media object. Since the invisible watermark can not be detected by the human eye we need some type of extraction algorithm to be able to read the watermark. Invisible watermarks do not change the signal to a perceptually great extent, i.e. [10] there are only minor variations in the output signal. An example of an invisible watermark is when some bits are added to an image modifying only its least significant bits. The example in the figure 3 shows the invisibly watermarked image.

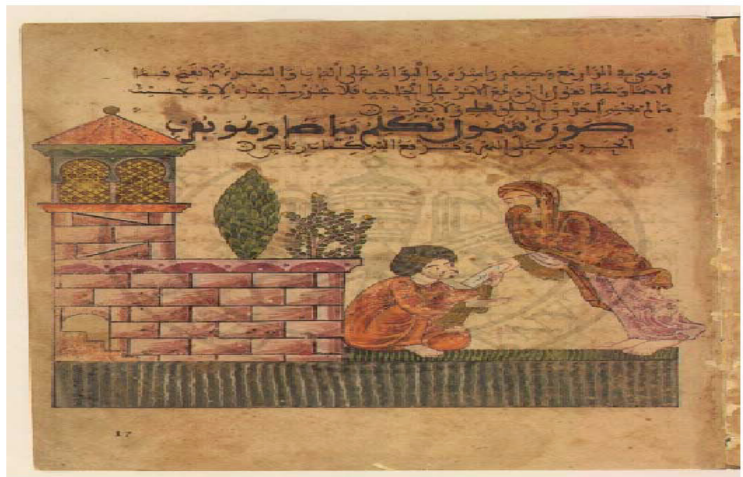

Figure – 2 Visible Watermarking

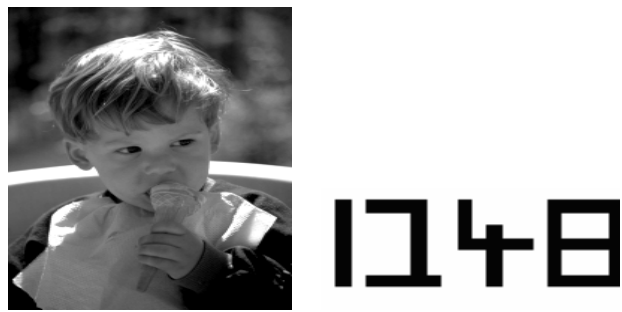

Figure – 3 Invisible Watermarking

# 5. CLASSIFICATION OF WATERMARKING TECHNIQUE

Watermarking techniques can be classified according to the application domain, according to the type of document according to the human perception and according to the application. Classification of watermarking techniques is shown in figure 4 Watermarks can be embedded into the multimedia content in spatial domain or in the frequency domain. Frequency domain watermarking methods may use several different domains, such as discrete cosine transformation (DCT) domain, [11] discrete Fourier transformation (DFT) domain, discrete wavelet transformation (DWT) domain, fast hadamard transform (FHT) domain etc. In the literature, it has been affirmed that the frequency domain techniques are more robust than spatial domain techniques. The watermarking algorithms can be named according to the embedded multimedia content, such as text, image, audio and video watermarking.

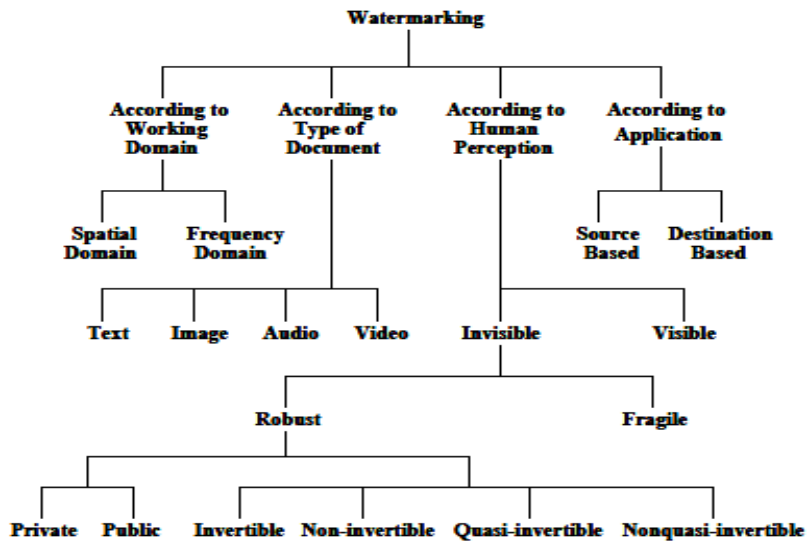

Figure - 4 Classification of Watermarking Technique

# 6. APPLICATION OF DIGITAL WATERMARKING

Digital watermarking technologies have been proposed to be implemented in many applications .Instead of compiling an exhaustive list of digital watermarking application; Following are the common applications of watermarking.

## 6.1 Copyright Protection

Copyright protection is an important application of digital watermarking. It enables the identification of the copyright holder and thus protects his or her right in content distribution. Watermarking is embedded into an image to protect the rights of the owner. It should be possible to detect the watermark despite common image processing, geometrical distortions, image compression and many other types of image manipulations.

## 6.2 Authentication

In order to be able to validate the content, any change to or manipulation with the content should be detected. This can be achieved through the use of "fragile/semi-fragile watermark" which has the low robustness to the modifications of the host image. The semi-fragile watermarking can also serve the purpose of quality measurement. The extracted watermark can not only notify the possible tampering with the host image, but also provide more information about the degradation of the host image, such as Peak Signal to Noise Ratio (PSNR) of the degraded host image. This can be very useful for broadcasting or network transmission, since sometimes the original reference is not available at the receiver side.

## 6. 3 Temper Detection and Localization

Temper detection is used to disclose alterations made into an image. It is closely related to authentication. If tempering is detected in an image, then the image is considered unauthentic. Temper localization enables further investigation of an act of tempering by identifying the tempered region within the image. This information can assist in media forensics.

## 6. 4 Media Forensics

In 2006 Fridrich. J., J. Lukas and M.Goljan [12] describes a Media forensics involves the investigation of digital data in order to unveil scientifically valid information for court evidence [13]. The deleted and hidden data are usually discovered using digital tool. The applications of digital watermarks in media forensics include the trustworthy digital camera, traitor tracing, transaction tracking and content recovery.

## 6. 5 Data Monitoring

Data monitoring is a way to save data of what is transmitted from for example a TV-channel. In 1997 it was discovered that some TV stations in Japan where over booking their air time for commercials. They got paid by advertisers for hours of commercials that were never aired [11]. Watermarking can be used as part of an automatic monitoring system that stores information about what have been aired by the TV stations. Data monitoring can also be used for statistical data collection and analysis [14].

## 6. 6 Time Stamping

Time stamping with the use of watermarking will make it possible for media objects contain information about when they were created or last used [15]. The time stamp can be used for copyright protection but also in other instances when it is important that the media has a time mark. For example security cameras will need to have an exact time for the images being recorded.

## 6. 7 Copy and Usage Control.

Users can have different privilege (play/copy control) on the object due to different payment for that object. It is expected in some systems to have a copy and usage control mechanism to check illegal copies of the content or limit the number of times of copying. A watermark can be used for this purpose.

## 6. 8 System Enhancement

This type of applications, where a device is designed to react to watermark for the benefit of the user, is also referred to as device control applications [16]. More recently for example Philips and Microsoft have demonstrated an audio watermarking system for music. Basically, as music is played, a microphone on a PDA can capture and digitize the signal, extract the embedded watermark and based on information encoded in it, identify the song. If a PDA is network connected, the system can link to a database and provide some additional information about the song, including information about how to purchase it.

# 7. PROPERTIES OF WATERMARKING

The main properties of the watermarks are robustness, fidelity, computational cost, data Payload, and false positive rate [11].On the other hand, one property may confront with another. Increasing the strength of the watermark can increase the robustness but it decreases the fidelity [17]. There must be a trade-off between the requirements and properties of the watermarking schemes depending on the applications. Under this following are the common properties of watermarking.

## 7.1 Robustness

A robust watermark must be invariant to possible attacks and remains detectable after attacks are applied. However, it is probably impossible, up to now, for a watermark to resist all kinds of attacks, in addition, it is unnecessary and extreme. The robustness criterion is specific for the type of application. On the other hand, the concept of fragile watermarks conflicts with the robustness criteria. In these applications, the watermark must be changed or lost after any applied attack. In many applications, when the signal processing between embedding and detection is unpredictable, the watermark may need to be robust to every possible distortion.

## 7. 2 Fidelity

Fidelity is a major concern for invisible types of watermarks. High fidelity means that, the amount of degradation caused by the watermark in the quality of the cover image is unnoticeable to the viewer. Nevertheless, it is Possible for the watermarked work to be degraded in the transmission process earlier until it being seen by a person, a different definition of fidelity might be further suitable. [17].

## 7. 3 Computational cost

The speed of the watermark embedding operation is a very important issue especially in broadcast monitoring applications where it must not slow down the media production and the watermark detector must work in real-time while monitoring the broadcasts

## 7. 4 Data Payload

The Data payload defines the number of bits a watermark embeds in a single unit of time. Diverse applications demand diverse data payload. For instance, Copy control applications may require a few bits embedded in cover works.

## 7. 5 False Positive Rate

A false positive is the identification of a watermark from a covered work which does not contain one in reality. The false positive rate is the probability of falsely rejecting the null hypothesis for a particular test among all the tests performed. If the false positive rate is a constant α for all tests performed, it can also be interpreted as the expected proportion among all tests performed that are false positives.

# 8. IMPLEMENTATION AND RESULTS

## 8.1 Working Of The Edge Detection Using Gabor Filters

The Gabor filter was originally introduced by Dennis Gabor since 1946 [18]. Gabor filters have been successfully applied to many image-processing applications, such as texture segmentation, [19] document analysis, edge detection, retina identification, fingerprint processing, image coding, and image representation. An advantage of these filters is that they satisfy the minimum Space- bandwidth product per the uncertainty principle [20]. Gabor filters have the ability to the minimum space bandwidth product per the uncertainty principle. They provide simultaneous optimal resolution in both the space and spatial-frequency domains.Gabor filters are directly related to Gabor wavelets they can be designed for a dilations and rotations. The Gabor space is very useful in e.g., image processing applications such as iris recognition and fingerprint recognition. A Gabor filter can be viewed as a sinusoidal plane of a particular frequency and orientation, modulated by a Gaussian envelope [21]. Gabor filter is a linear filter that is created by modulating a sinusoid with a Gaussian. This filter is given by

$$g(x, y; \lambda, \theta, \phi, \sigma, \gamma) = e^{\left(-\frac{x'^2 + \gamma^2 y'^2}{\sigma^2}\right)} \cos\left(2\pi\frac{x'}{\lambda} + \phi\right)$$

Where

$$x' = x\cos\theta + y\sin\theta$$

And

$$y' = -x\sin\theta + y\cos\theta$$

The parameters used in the above equation for g are explained below

1. $\sigma$ is the standard deviation of the Gaussian factor and determines the (linear) size of its receptive field.

2. Cosine factor of the Gabor filter is $\lambda$ wavelength.

3. $\theta$ specifies the orientation of the normal to the parallel stripes of the Gabor filter.

4. $\phi$ is the phase offset of the cosine factor and determines the symmetry of the Gabor filter.

5. Ellipticity of the Gaussian factor is called $\Upsilon$ spatial aspect ratio.

We also define another parameter called the Bandwidth (b) of a Gabor filter. The half-response spatial frequency bandwidth of a Gabor filter is related to the ratio $\sigma/\lambda$ where $\sigma$ and $\lambda$ are the standard deviation of the Gaussian factor of the Gabor function and the preferred wavelength, respectively as follows

$$b = \log_2 \frac{\frac{\sigma}{\lambda}\pi + \sqrt{\frac{\ln 2}{2}}}{\frac{\sigma}{\lambda}\pi - \sqrt{\frac{\ln 2}{2}}}$$

Gabor filters can also be used for edge recognition. This process involves firstly super positioning different Gabor filters which are at different phases and orientations, and secondly performing a convolution of the filters with the original image. What's new for this function is that accepts multiple values for φ, and it also accepts the number of oscillations. The multiple φ values are used to build multiple Gabor filters with different phases. The number of oscillations tells the function to construct the specified number Gabor functions with different orientations θ, evenly distributed from θ= 0° to θ= 360°. These Gabor filters are then superimposed, and the superimposed filter is convoluted with the real and imaginary parts of the image matrix. The final result of the edge detection is obtained by calculating the absolute values of the results of the convolution. figure 6 shows the result of applying the Gabor edge detection algorithm to a high resolution version of figure 5. This operation was carried out with λ=8, θ= 10°, φ= {0°, 90°}, ϒ = 0.5° bandwidth = 1, and number of oscillations = 12.

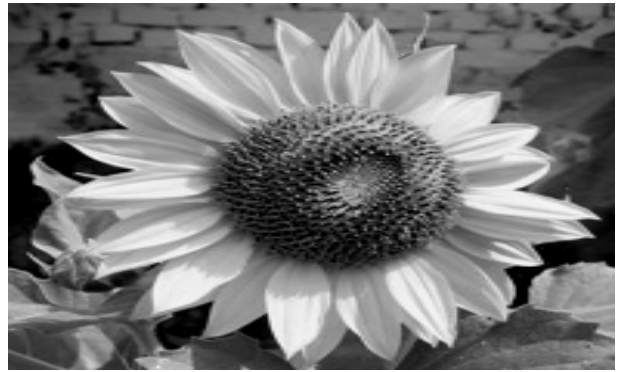

Figure - 5 The Original image of Sunflower

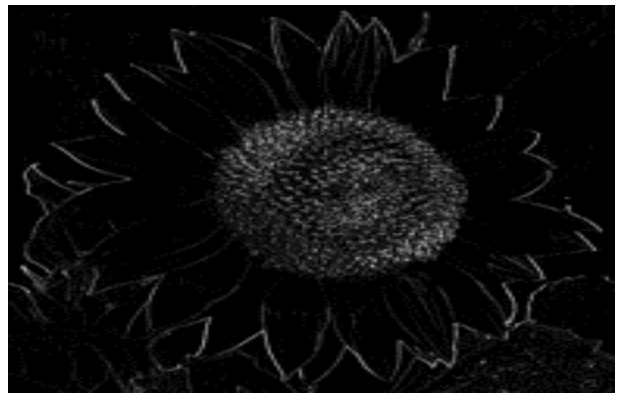

Figure - 6 The Result of after applying the Gabor edge detection algorithm

## 8. 2 We Are Using Data Hiding By Least Significant Bit Substitution Algorithm

This method embeds the fixed-length secret bits in the same fixed length LSB of pixels. Least Significant Bit (LSB) substitution method is a very popular way of embedding secret messages with simplicity. The fundamental idea here is to insert the secret message in the least significant bits of the images. This actually works because the human visual system is not sensitive enough to pick out changes in color whereas changes in luminance are much better picked out. A basic algorithm for LSB substitution is to take the first N cover pixels where N is the total length of the secret message that is to be embedded in bits. After that every pixels last bit will be replaced by one of the message bits [22]. As an example, suppose that we have two adjacent pixels (six bytes) with the following RGB encoding.

10110101  01001101  11001101
00010011  00010100  01001010

Now suppose we want to hide the following 6 bits of data 001001. If we overlay these 6 bits over the LSB of the 6 bytes above, we get the following, where bits in **bold** have been changed.

1011010**0**  0100110**0**  11001101
0001001**0**  00010100  0100101**1**

We have successfully hidden 6 bits but only at the cost of changing 4, (roughly 66.75%), of them LSB. This paper concerns with hiding text data containing a secret message within images with the Least Significant Bit (LSB) substitution procedure. Therefore, a general and common model for evaluating the entire process is required. The procedure should include all the steps relevant with the desired message hiding within images. This process must begin with gathering images that are to be operated on. Further, the images have to produce relevant and key variables that will significantly describe the environment. Two of the steps are the most important phases of Least Significant Bit (LSB) substitution procedure. They are embedding and extraction of the secret text message. Because with these two processes, the main function of embedding the data into images and further extraction will be done.

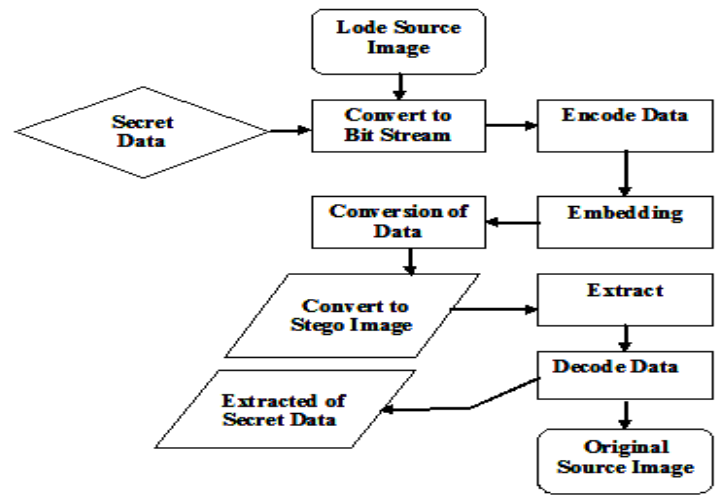

Figure – 7 Least Significant Bit (LSB) Substitution Procedure

In this the paper is concerned with images, the source can be any satellite image or any other images taken from digital camera or scanner. After that gathering image these collected images are stored on data disk as well as USB. After the source image is located, the embedding and extraction will be done and finally the receiver will be able to enjoy the original source image.

Firstly is acquiring the source image file and then performs the necessary calculations to estimate the pixel values of red, green and blue color in the image. This will be followed by performing filtering process which will remove any noisy information from the original data. After that embed data within the source image file, the first task is to choose the relevant variables that effectively describe the environment because it is necessary to identify the specific attributes and characteristics of the source image in terms of specifying the pixel values of red, green and blue colors. The source image can be captured by using a digital camera or scanner. The secret data that are to be hidden can be in any kind of digital form. The text containing the data to be hidden will be chosen by the source image sender. There is a specific type of image that is used. That is the source image which the secret data is to be embedded and password will be given by the sender in order to keep the data safe and to avoid access from malicious users.

In this paper we will perform the data hiding by using Least Significant Bit substitution algorithm. The source image components have been examined in terms of splitting the RGB components into 8- bit binary streams. In this step, the Least Significant Bit (LSB) of every pixel octet will be replaced by the secret message which have already transformed into 8-bit binary streams. This substitution will be performed in character by character manner and tricky part of this substitution is that here an encryption key will be used which will be used as the security key for the correct recognition in the receiver portion side. The key will be given by the source sender side. In our experimental results, we have three gray scale images which are shown in figure 9 were used as cover images. The characters from the string are converted into their corresponding binary equivalent of ASCII code. The bits of character are then embedded into three source image pixels around a lead pixel per character. The watermark extraction stage the key information required is the pixel value used as a key, around which the watermark bits are placed. The figure 9 shows various images (a1, a2, a3) upon which the algorithm was implemented and their corresponding watermarked copy (b1, b2, b3). The values for the mean square error (MSE) and peak signal to noise ratio (PSNR) are measured. Table.1 lists all the experimental results obtained for various sized images. The quality of watermarked image can be evaluated with that of the real image using PSNR (Peak Signal to Noise Ratio) method. A higher PSNR ensures the watermarked image is not significantly distorted from the original.

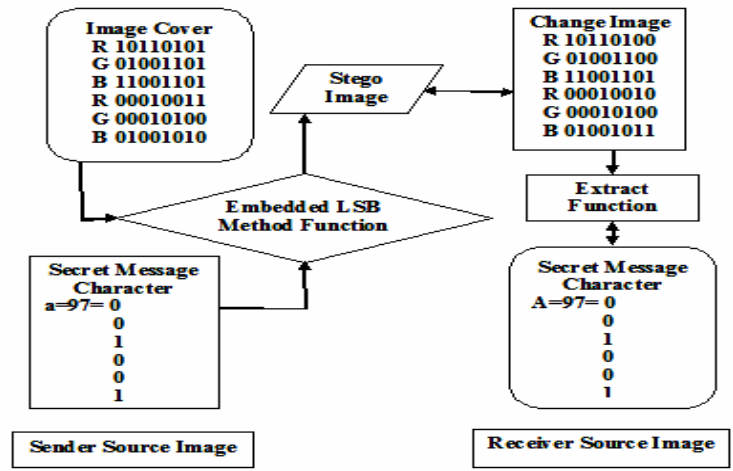

Figure – 8  The Pixels Fragmented into Bit

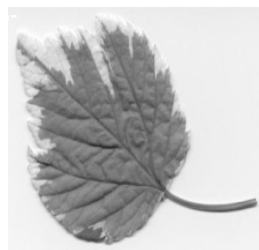　　　　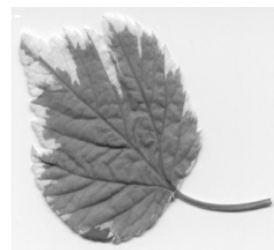

(a1) Original Image　　　　　　(b1) Watermarked Image

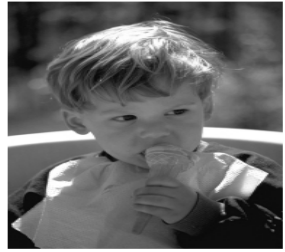 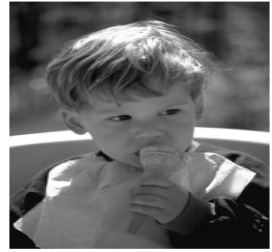

(a2) Original Image          (b2) Watermarked Image

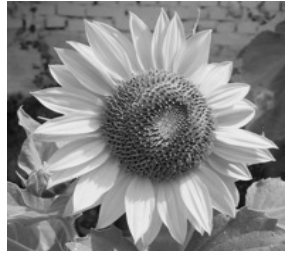 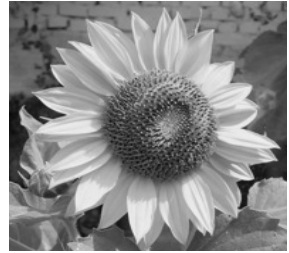

(a3) Original Image          (b3) Watermarked Image

Figure – 9 The Watermark Embedding Process Original image (a1, a2, a3) Watermarked Image (b1, b2, b3)

| Image Name | Image Size | Number Of Characters Embedded | Number Of Bits Embedded | Peak Signal to Noise Ratio (PSNR) dB | Mean Square Error (MSE) |
|------------|-----------|-------------------------------|-------------------------|--------------------------------------|-------------------------|
| Leaf | 256*240 | 864 | 5342 | 56. 36 | 0. 09 |
| kid | 512*512 | 1248 | 8692 | 58. 42 | 0. 06 |
| Sunflower | 384*512 | 2680 | 18768 | 56. 22 | 0. 15 |

Table – 1 The Experimental result for MSE and PSNR   measurement

## 8. 3 Working With Algorithm Discrete Cosine Transform (DCT) For Based On Robustness

Discrete-Cosine-Transform or DCT is a popular transform domain watermarking technique. The DCT permission image to be split into different frequency bands example for  the high, middle and low frequency bands and making it simple way to select the band in which the watermark is to be inserted. The DCT based on Mid Band Exchange Coefficient (MBEC) algorithm. DCT-based methods divide image into 8×8 sized blocks and then transforms images of size N×N into the DCT coefficient matrix with the same size . MBEC use the one bit of binary watermark image. The MBEC watermarking algorithm encodes one-bit of the binary watermark image into one 8x8 DCT sub-block of the original image. If the difference of two mid-band coefficients is positive in the case of the encoded value is 1 means the first coefficient is small then second coefficient then we encoded value is 1. Otherwise, these two mid-band coefficients are exchanged.

A classical middle band based algorithm is quite robust against JPEG compression and common image manipulation operations. The basic idea of the classical MBEC scheme was discussed in [23]. In 8x8 DCT block the middle-band frequency region is denoted by FM. The low frequency component of the block is denoted by FL, while FH is used to denote the higher frequency components. The FM is selected as the embedding region as to provide additional resistance to lossy compression. Next, from the FM region, two locations Pi (u1, v1) and Pi (u2, v2) are chosen in the ith DCT block for comparison. From the table 2 the coefficients at (5, 2) and (4, 3) have value 22 or (2, 3) and (4, 1) have value 14, would make suitable candidates for comparison, as their quantization values are equal. The DCT block will encode a "0" if the value of the first pixel position is greater than or equal to the second pixel position otherwise it will encode a "1". In other word if Pi (u1, v1) ≥ Pi (u2, v2) then DCT block a value "0" and if Pi (u1, v1) < Pi (u2, v2) it will encode a "1".

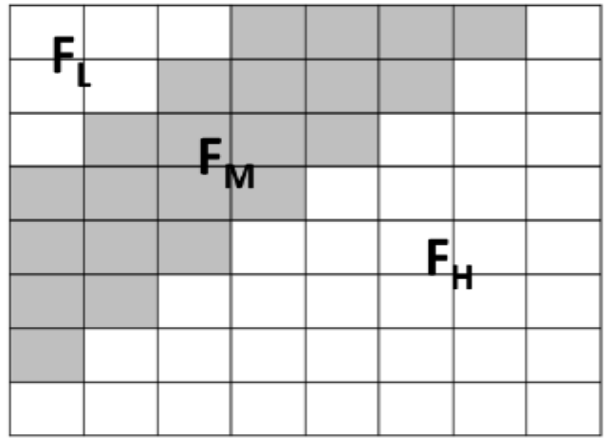

Figure - 10 Definition of DCT Region

The equations for two dimensional discrete cosine transform (2D-DCT) is defined as

$$F(jk) = a(j)a(k)\sum_{m=0}^{N-1}\sum_{n=0}^{N-1} f(mn)cos\left[\frac{(2m+1)j\pi}{2N}\right]cos\left[\frac{(2n+1)k\pi}{2N}\right]$$

| 16 | 11 | 10 | 16 | 24 | 40 | 51 | 61 |
|----|----|----|----|----|----|----|----|
| 12 | 12 | 14 | 19 | 26 | 48 | 16 | 55 |
| 14 | 13 | 16 | 24 | 40 | 57 | 69 | 56 |
| 14 | 17 | 22 | 29 | 51 | 87 | 80 | 62 |
| 18 | 22 | 37 | 56 | 68 | 109 | 108 | 77 |
| 24 | 35 | 55 | 64 | 81 | 104 | 113 | 92 |
| 49 | 64 | 78 | 87 | 103 | 121 | 120 | 101 |
| 72 | 92 | 95 | 98 | 112 | 100 | 103 | 99 |

Table – 2 Quantization values used in transforms Domain

The corresponding inverse transformation two dimensional (2D - IDCT) is defined as

$$f(mn) = \sum_{m=0}^{N-1}\sum_{n=0}^{N-1} a(j)a(k)F(jk)cos\left[\frac{(2m+1)j\pi}{2N}\right]cos\left[\frac{(2n+1)k\pi}{2N}\right]$$

The algorithm which is used to embed a watermark on an image is given below. We take input as Original Image and Watermark data and produce output as Watermarked Image.

1) Fragment the image into non-overlapping blocks of 8x8.

2) Using DCT to each of these blocks.

3) Apply some block selection criteria based on the knowledge of Human Visual System (HVS)

4) Apply coefficient selection criteria (e.g. highest, mid, lowest).

5) Embed watermarks by modifying the selected coefficients.

6) Apply inverse DCT transforms on each block.

Two metrics for quality of watermarked images have been used which are Peak Signal to Noise Ratio (PSNR). This Peak Signal to Noise (PSNR) is defined as

$$PSNR = 10log_{10}\frac{A^2}{\frac{1}{N \times M}\sum_{i=1}^{N}\sum_{j=1}^{M}[f(i,\ j) - f'(i,\ j)]^2}$$

Another the comparability of the distilled watermark with the original watermark is quantitatively analysis by using Normalized Cross-Correlation (NC) [24]. The Normalized Cross-Correlation (NC) is defined as

$$NC = \frac{\sum\limits_{i=1}^{M_1}\sum\limits_{j=1}^{M_2} W(i, j) \cdot W'(i, j)}{\sqrt{\sum\limits_{i=1}^{M_1}\sum\limits_{j=1}^{M_2}[W(i, j)]^2}\sqrt{\sum\limits_{i=1}^{M_1}\sum\limits_{j=1}^{M_2}[W'(i, j)]^2}}$$

The Normalized Cross-Correlation (NC) has value [0, 1] calculated using the following equation. If NC = 1 then the embedded watermark and the extracted watermark are same. Generally the value of NC>0.7500 is accepted as reasonable watermark extraction. Where W is Original Watermark and W′ is detected watermark. Its unit is db and the bigger the PSNR value is better the watermark conceals [25].The experiment is simulated with the software MATLAB. In the following experiments, the gray-level image Leaf, Kid and Sunflower is used as host image to embed watermark. The watermark transposed is shown in figure 11.

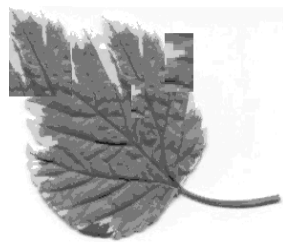 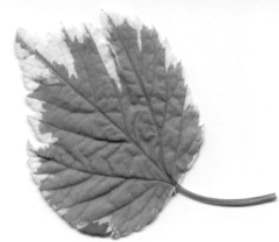

(a1) Transposed Image          (b1) Watermarked Image

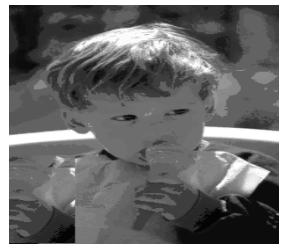 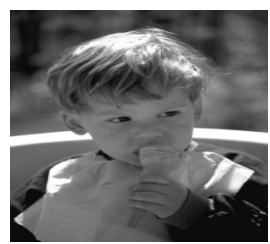

(a2) Transposed Image          (b2) Watermarked Image

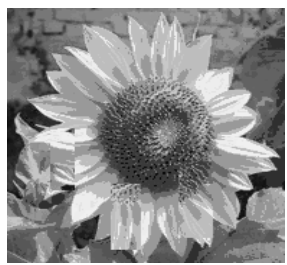 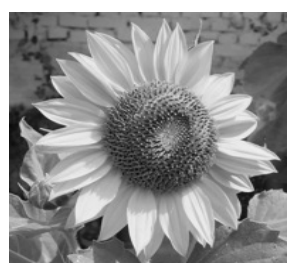

(a3) Transposed Image          (b3) Watermarked Image

Figure – 11 The Watermark Transposed image (a1, a2, a3) and Watermarked Image (b1, b2, b3)

The following Table 3 shows the PSNR of the different watermarked images and the Normalized Cross-Correlation (NC) of their extract watermarks. When PSNR is higher than 35 dB Watermarked image has a very good quality and the eye could hardly tell the difference between the original and the Watermarked image. While when NC is higher than 0.7500 the extracted

Watermarked is considered as valid one. From the below Table 3 we can safely say that the watermarking schema discuss in this paper has a good invisibility and can extract the marks correct.

| Image Name | Number Of Bits Embedded | Normalized Cross Correlation (NC) | Peak Signal to Noise Ratio (PSNR) dB |
|------------|------------------------|-----------------------------------|--------------------------------------|
| Leaf | 5342 | 0. 8342 | 39. 5607 |
| kid | 8692 | 0. 9682 | 48. 3421 |
| Sunflower | 18768 | 0. 8768 | 41. 2722 |

Table – 3 PSNR and NC values for different images

## 12. CONCLUSION

Digital Watermarking defines methods and technologies that hide information, for example a number or text, in digital media, such as images, copyright protection, tamper proofing, video or audio. In this paper we are briefly defining the concepts of watermarking, history of watermarks and the properties of a watermarking system as well as an application. We are proposing edge detection from Gabor Filter method. In this paper we are using data hiding by the simple LSB substitution method. In the method a set of pixels that constitute a block jointly share the bits from the watermark .The values for the mean square error (MSE) and peak signal to noise ratio (PSNR) are measured. The results indicate the method introduces low noise and hence ensures lesser visible distortions. The PSNR value of about 58 dB attained is much higher when compared with any other proposed methods. We have successfully implemented the LSB method and got satisfactory results as we also took into account the filtering option for making sure that the image is noise-free before transmission. The techniques in the spatial domain still have relatively low-bit capacity. The frequency domain based techniques can embed more bits for watermark and are more robust to attack. In this paper, we use DCT transform watermark algorithms based on robustness. The hardiness of the watermarking evaluated have been measured by the Peak Signal to Noise Ratio (PSNR) of the watermarked Image and similarity between real Watermark and after extract Watermark using Normalized Cross Correlation (NC) method. In this paper Implementation results show that the imperceptibility of the watermarked image is acceptable.

## REFERENCES

[1]   Anderson, R.J. & Petitcolas, F., (1998) "On the Limits of Steganography", IEEE Journal of Selected Areas in Communications, vol 16, Issue 4, pp. 474 – 481

[2]   Hartung F.  and Kutter M. ,(1999) "Multimedia Watermarking Techniques", Proc. IEEE, vol 87, no.7, pp 1079-1107

[3]   Marvel, L. M., Boncelet Jr. , C.G. , Retter, C. , (1999)  "Spread Spectrum Steganography",   IEEE Transactions on image processing

[4]  Bender W. and Morimoto N., Lu A., (1996) "Techniques for data hiding", IBM system journal , vol 35, no 3 & 4, pp 313 – 336

[5]  Kutter M. & Hartung F., (1999) "Introduction to Watermarking Technique in Information Techniques for Steganography and Digital Watermarking", S.C. Katzenbeisser et al., Eds. Northwood, MA: Artec House, pp 97-119

[6]  Provos, N. & Honeyman, P.,(2003) "Hide and Seek: An introduction to steganography", IEEE Security and Privacy Journal, vol 1 Issue 3 , pp 32 - 44

[7]  Hartung F. and Kutter M., (1999) "Multimedia watermarking techniques", In Proceedings IEEE: Special Issue on Identification and Protection of Multimedia Information, vol 87(7), pp 1079–1107

[8]  Cox Ingemar J. , Kilian Joe, Leighton Tom, Shamoon Talal, (1995) "Secure spread spectrum watermarking for multimedia", NEC Research Institute, Technical Report

[9]  Kutter M. and Petitcolas F. A. P. ,"A fair benchmark for image watermarking systems",(1999) In Proceedings of the SPIE Security and Watermarking of Multimedia Contents, vol 3657, pp 226 - 239

[10] Mohanty S. P., Ramakrishnan K. R., and Kankanhalli M., (1999) "A dual watermarking technique for images", In Proceedings of the 7th ACM International Multimedia Conference, ACM-MM'99, pp 49 - 51

[11] Ingemar Cox J., Matt L. Miller, Jeffrey A. Bloom, (2000)"Watermarking applications and their properties", in International Conference on Information Technology, ITCC 2000, pp 6-10

[12] Fridrich. J. , Lukas J. and Goljan M . , (2006) "Digital Camera Identification from Sensor Noise", IEEE Transactions on Information Security and Forensics, vol 1(2),  pp 205-214

[13] Allgeier. M. (2005) "Digital Media Forensics", online at ,http://www.securityfocus.com/infocus/1253.

[14] Kalker T., Depovere G., Haitsma J., and Maes M.,(1999) "A video watermarking system for broadcast monitoring", In Proceedings SPIE Electronic Imaging , Security and Watermarking of Multimedia Contents, San Jose, CA, pp 103-112

[15] Chang C. C. , Hwang K. F. , and Hwang M. S.,(2002) "Robust authentication scheme for protecting copyrights of images and graphics", IEE Proceedings Vision, Image and Signal Processing, vol 149, no 1,  pp 43–50

[16] Cox, I. J., Miller M . L., Bloom, J.A., (2001) "Digital Watermarking," Morgan Kaufmann  Publishers, San Francisco

[17] Katzenbeisser S. and Petitcolas F. A. P., (2000) "Information Hiding Techniques for Steganography and Digital Watermarking", Artech House, UK

[18] Gabor D., (1946) "Theory of communication", Journal of the Institute of Electrical Engineers, vol 93 , pp 429 – 457

[19] Porat M. and Zeevi Y. , (1988)"The generalized Gabor scheme of image representation in biological and machine vision" , IEEE Trans. Pattern Anal. Machine Intell., vol 10, pp 452- 468, doi 10.1109/34.3910

[20] Dunn D. F.  and Higgins W. E. , (1995) "Optimal Gabor filters for texture segmentation" , IEEE Trans. Image Proc., vol 4, pp 947- 964

[21] Movellan Javier R., (2002) "Tutorials on Gabor Filters", GNU Free documentation License 1.1, Kolmogorv Project, pp 1-20

[22] Wang R.Z. , Lin C.F. , Lin J.C. , (2001)"Image hiding by optimal LSB substitution and genetic algorithm", Pattern Recognition, vol 34 , no 3 , pp 671 – 683

[23] Liu L. S. , Li R. H., Gao Q. ,(2004) "Method of embedding digital watermark into the green component of color image," In Journal of XianJiaotong university, vol 38, pp 1256-1259

[24] Jiang J. and Armstrong A. ,(2002) "Data hiding approach for efficient image indexing", Electronics letters. 7th, vol 38, no 23, pp 1424 - 1425

[25] Chandra Mohan B. , Veera Swamy K. and Srinivas Kumar S. , (2011) "A Comparative performance evaluation of SVD and Schur Decompositions for Image Watermarking ", IJCA Proceedings on International Conference on VLSI, Communications and Instrumentation (ICVCI) (14), pp 25–29